%% file: RTN-120.tex
\documentclass[]{spie}

\usepackage[nonumberlist,nogroupskip,toc,numberedsection=autolabel,style=index]{glossaries}

\usepackage{amsmath,amsfonts,amssymb}
\usepackage{graphicx}
\usepackage{listings}
\usepackage{xcolor}
\usepackage[colorlinks=true, allcolors=blue, backref=page]{hyperref}
\usepackage{xspace}
\usepackage{listings}


\definecolor{codegreen}{rgb}{0,0.6,0}
\definecolor{codegray}{rgb}{0.5,0.5,0.5}
\definecolor{codepurple}{rgb}{0.58,0,0.82}
\definecolor{backcolour}{rgb}{0.95,0.95,0.92}

\lstdefinestyle{rtnstyle}{
    backgroundcolor=\color{backcolour},
    commentstyle=\color{codegreen},
    keywordstyle=\color{magenta},
    numberstyle=\tiny\color{codegray},
    stringstyle=\color{codepurple},
    basicstyle=\ttfamily\footnotesize,
    breakatwhitespace=false,
    breaklines=true,
    captionpos=b,
    keepspaces=true,
    numbers=none,
    numbersep=5pt,
    showspaces=false,
    showstringspaces=false,
    showtabs=false,
    tabsize=2,
    frame=single,
    rulecolor=\color{backcolour},
    framexleftmargin=6pt,
    framexrightmargin=6pt,
    framextopmargin=4pt,
    framexbottommargin=4pt
}
\newcommand\YAMLcolonstyle{\color{black}\mdseries}
\newcommand\YAMLkeystyle{\color{magenta}\mdseries}
\newcommand\YAMLvaluestyle{\color{codepurple}\mdseries}

\lstdefinelanguage{YAML}{
    keywords={true,false,null,y,n},
    keywordstyle=\color{darkgray}\bfseries,
    basicstyle=\YAMLkeystyle,                                 
    sensitive=false,
    comment=[l]{\#},
    morecomment=[s]{/*}{*/},
    commentstyle=\color{black}\ttfamily,
    stringstyle=\YAMLvaluestyle\ttfamily,
    moredelim=[l][\color{black}]{\&},
    moredelim=[l][\color{magenta}]{*},
    moredelim=**[il][\YAMLcolonstyle{:}\YAMLvaluestyle]{:},   
    morestring=[b]',
    morestring=[b]",
    literate =    {---}{{\ProcessThreeDashes}}3
    {>}{{\textcolor{black}\textgreater}}1
    {|}{{\textcolor{black}\textbar}}1
    {\ -\ }{{\mdseries\ -\ }}3,
}

\usepackage[colorinlistoftodos]{todonotes}
\let\Oldtodo\todo
\renewcommand{\todo}[1]{\Oldtodo[inline, color=red]{#1}}

\providecommand{\secref}[1]{\hyperref[#1]{\S\ref{#1}}}
\providecommand{\appref}[1]{\hyperref[#1]{Appendix \ref{#1}}}
\providecommand{\tabref}[1]{\hyperref[#1]{\autoref{#1}}}
\providecommand{\figref}[1]{\hyperref[#1]{\autoref{#1}}}
\providecommand{\eqnref}[1]{\hyperref[#1]{Eq.~\ref{#1}}}
\providecommand{\recref}[1]{\hyperref[#1]{REC-\ref{#1}}}

\graphicspath{{figures}}

\input{aglossary}
\makeglossaries

\begin{document}
\input{authors}
\authorinfo{Corresponding author: Leanne P. Guy, leanne.guy@noirlab.edu}
\title{Development of a Retrieval-Augmented Generation Virtual Assistant for Enhanced Information Discovery at Rubin Observatory}

\hypersetup{
    pdftitle={Development of a Retrieval-Augmented Generation Virtual Assistant for Enhanced Information Discovery at Rubin Observatory},
    pdfauthor={guylp},
    pdfkeywords={}
}

\maketitle

\input{abstract}

\input{body}

\acknowledgments

This material is based upon work supported in part by the National Science Foundation through Cooperative Agreements AST-1258333 and AST-2241526 and Cooperative Support Agreements AST-1202910 and AST-2211468 managed by the Association of Universities for Research in Astronomy (AURA), and the Department of Energy under Contract No.\ DE-AC02-76SF00515 with the SLAC National Accelerator Laboratory managed by Stanford University.
Additional Rubin Observatory funding comes from private donations, grants to universities, and in-kind support from LSST-DA Institutional Members.

ADACS is funded from the Astronomy National Collaborative Research Infrastructure Strategy (NCRIS) allocation provided by the Australian Government and managed by Astronomy Australia Limited (AAL).

\bibliographystyle{spiebib}
\bibliography{local,lsst,lsst-dm,refs_ads,refs,books}

\printglossaries

\end{document}

%% file: aglossary.tex
\newacronym{AAL} {AAL} {Astronomy Australia Limited}
\newacronym{ADQL} {ADQL} {Astronomical Data Query Language (IVOA standard)}
\newacronym{AGN} {AGN} {Active Galactic Nuclei}
\newacronym{AI} {AI} {Artificial Intelligence}
\newacronym{API} {API} {Application Programming Interface}
\newacronym{AST} {AST} {NSF Division of Astronomical Sciences}
\newacronym{AURA} {AURA} {Association of Universities for Research in Astronomy}
\newacronym{AZ} {AZ} {Azimuth}
\newglossaryentry{Alert} {name={Alert}, description={A packet of information for each source detected with signal-to-noise ratio > 5 in a difference image by Alert Production, containing measurement and characterization parameters based on the past 12 months of LSST observations plus small cutouts of the single-visit, template, and difference images, distributed via the internet}}
\newglossaryentry{Alert Production} {name={Alert Production}, description={Executing on the Prompt Processing system, the Alert Production payload processes and calibrates incoming images, performs Difference Image Analysis to identify DIASources and DIAObjects, and then packages the resulting alerts for distribution}}
\newglossaryentry{Archive} {name={Archive}, description={The repository for documents required by the NSF to be kept. These include documents related to design and development, construction, integration, test, and operations of the LSST observatory system. The archive is maintained using the enterprise content management system DocuShare, which is accessible through a link on the project website www.project.lsst.org}}
\newglossaryentry{Archive Center} {name={Archive Center}, description={Part of the LSST Data Management System, the LSST archive center is a data center at NCSA that hosts the LSST Archive, which includes released science data and metadata, observatory and engineering data, and supporting software such as the LSST Software Stack}}
\newglossaryentry{Association Pipeline} {name={Association Pipeline}, description={An application that matches detected Sources or DIASources or generated Objects to an existing catalog of Objects, producing a (possibly many-to-many) set of associations and a list of unassociated inputs. Association Pipelines are used in Alert Production after DIASource generation and in the final stages of Data Release processing to ensure continuity of Object identifiers}}
\newglossaryentry{Association of Universities for Research in Astronomy} {name={Association of Universities for Research in Astronomy}, description={ consortium of US institutions and international affiliates that operates world-class astronomical observatories, AURA is the legal entity responsible for managing what it calls independent operating Centers, including LSST, under respective cooperative agreements with the National Science Foundation. AURA assumes fiducial responsibility for the funds provided through those cooperative agreements. AURA also is the legal owner of the AURA Observatory properties in Chile}}
\newacronym{BCE} {BCE} {Before Common Era}
\newglossaryentry{Butler} {name={Butler}, description={A middleware component for persisting and retrieving image datasets (raw or processed), calibration reference data, and catalogs}}
\newacronym{CA} {CA} {Certificate Authority}
\newacronym{CD} {CD} {Critical Design}
\newglossaryentry{Center} {name={Center}, description={An entity managed by AURA that is responsible for execution of a federally funded project}}
\newglossaryentry{Commissioning} {name={Commissioning}, description={A two-year phase at the end of the Construction project during which a technical team a) integrates the various technical components of the three subsystems; b) shows their compliance with ICDs and system-level requirements as detailed in the LSST Observatory System Specifications document (OSS, LSE-30); and c) performs science verification to show compliance with the survey performance specifications as detailed in the LSST Science Requirements Document (SRD, LPM-17)}}
\newglossaryentry{Construction} {name={Construction}, description={The period during which LSST observatory facilities, components, hardware, and software are built, tested, integrated, and commissioned. Construction follows design and development and precedes operations. The LSST construction phase is funded through the NSF MREFC account}}
\newglossaryentry{Contract} {name={Contract}, description={A binding legal agreement between parties obligating the one (typically the  'seller') to furnish certain supplies or services and the other (typically, the buyer) to compensate the seller for the supplies or services with some form of consideration, (typically money). The term, 'contract' is used interchangeably with 'sub-award' 'agreement' 'memorandum of understanding and/or agreement' and 'purchase order' Each is a term used to differentiate between a purchase-order-format type document and a complex purchase in a subcontract/sub-award-format type document. These also include awards and notices of awards; job orders or task letters issued under basic ordering agreements; letter contracts; orders, such as purchase orders and subcontracts under which the order becomes effective by written acceptance or performance; and bilateral contract modifications}}
\newacronym{DAC} {DAC} {Data Access Center}
\newacronym{DCR} {DCR} {Differential Chromatic Refraction}
\newacronym{DE-AC02} {DE-AC02} {Department of Energy contract number prefix}
\newacronym{DIA} {DIA} {Difference Image Analysis}
\newglossaryentry{DIAObject} {name={DIAObject}, description={A DIAObject is the association of DIASources, by coordinate, that have been detected with signal-to-noise ratio greater than 5 in at least one difference image. It is distinguished from a regular Object in that its brightness varies in time, and from a SSObject in that it is stationary (non-moving)}}
\newglossaryentry{DIASource} {name={DIASource}, description={A DIASource is a detection with signal-to-noise ratio greater than 5 in a difference image}}
\newacronym{DM} {DM} {Data Management}
\newacronym{DMS} {DMS} {Data Management Subsystem}
\newacronym{DOE} {DOE} {Department of Energy}
\newacronym{DP0} {DP0} {Data Preview 0}
\newacronym{DP1} {DP1} {Data Preview 1}
\newacronym{DP2} {DP2} {Data Preview 2}
\newacronym{DR} {DR} {Data Release}
\newacronym{DR1} {DR1} {Data Release 1}
\newacronym{DRP} {DRP} {Data Release Processing}
\newglossaryentry{Data Access Center} {name={Data Access Center}, description={Part of the LSST Data Management System, the US and Chilean DACs will provide authorized access to the released LSST data products, software such as the Science Platform, and computational resources for data analysis. The US DAC also includes a service for distributing bulk data on daily and annual (Data Release) timescales to partner institutions, collaborations, and LSST Education and Public Outreach (EPO). }}
\newglossaryentry{Data Management} {name={Data Management}, description={The LSST Subsystem responsible for the Data Management System (DMS), which will capture, store, catalog, and serve the LSST dataset to the scientific community and public. The DM team is responsible for the DMS architecture, applications, middleware, infrastructure, algorithms, and Observatory Network Design. DM is a distributed team working at LSST and partner institutions, with the DM Subsystem Manager located at LSST headquarters in Tucson}}
\newglossaryentry{Data Management Subsystem} {name={Data Management Subsystem}, description={The Data Management Subsystem is one of the four subsystems which constitute the LSST Construction Project. The Data Management Subsystem is responsible for developing and delivering the LSST Data Management System to the LSST Operations Project}}
\newglossaryentry{Data Management System} {name={Data Management System}, description={The computing infrastructure, middleware, and applications that process, store, and enable information extraction from the LSST dataset; the DMS will process peta-scale data volume, convert raw images into a faithful representation of the universe, and archive the results in a useful form. The infrastructure layer consists of the computing, storage, networking hardware, and system software. The middleware layer handles distributed processing, data access, user interface, and system operations services. The applications layer includes the data pipelines and the science data archives' products and services}}
\newglossaryentry{Data Release Processing} {name={Data Release Processing}, description={Deprecated term; see Data Release Production}}
\newglossaryentry{Data Release Production} {name={Data Release Production}, description={An episode of (re)processing all of the accumulated LSST images, during which all output DR data products are generated. These episodes are planned to occur annually during the LSST survey, and the processing will be executed at the Archive Center. This includes Difference Imaging Analysis, generating deep Coadd Images, Source detection and association, creating Object and Solar System Object catalogs, and related metadata}}
\newglossaryentry{Department of Energy} {name={Department of Energy}, description={cabinet department of the United States federal government; the DOE has assumed technical and financial responsibility for providing the LSST camera. The DOE's responsibilities are executed by a collaboration led by SLAC National Accelerator Laboratory}}
\newglossaryentry{Difference Image} {name={Difference Image}, description={Refers to the result formed from the pixel-by-pixel difference of two images of the sky, after warping to the same pixel grid, scaling to the same photometric response, matching to the same PSF shape, and applying a correction for Differential Chromatic Refraction. The pixels in a difference thus formed should be zero (apart from noise) except for sources that are new, or have changed in brightness or position. In the LSST context, the difference is generally taken between a visit image and template. }}
\newglossaryentry{Difference Image Analysis} {name={Difference Image Analysis}, description={The detection and characterization of sources in the Difference Image that are above a configurable threshold, done as part of Alert Generation Pipeline}}
\newglossaryentry{Differential Chromatic Refraction} {name={Differential Chromatic Refraction}, description={The refraction of incident light by Earth's atmosphere causes the apparent position of objects to be shifted, and the size of this shift depends on both the wavelength of the source and its airmass at the time of observation. DCR corrections are done as a part of DIA}}
\newglossaryentry{Docker} {name={Docker}, description={A system for packaging and distributing software using self-contained containers which may be run on any Linux system; \url{https://www.docker.com/}}}
\newglossaryentry{DocuShare} {name={DocuShare}, description={The trade name for the enterprise management software used by LSST to archive and manage documents}}
\newglossaryentry{Document} {name={Document}, description={Any object (in any application supported by DocuShare or design archives such as PDMWorks or GIT) that supports project management or records milestones and deliverables of the LSST Project}}
\newacronym{ECDFS} {ECDFS} {Extended Chandra Deep Field-South Survey}
\newacronym{EPO} {EPO} {Education and Public Outreach}
\newglossaryentry{Education and Public Outreach} {name={Education and Public Outreach}, description={The LSST subsystem responsible for the cyberinfrastructure, user interfaces, and outreach programs necessary to connect educators, planetaria, citizen scientists, amateur astronomers, and the general public to the transformative LSST dataset}}
\newacronym{FITS} {FITS} {Flexible Image Transport System}
\newglossaryentry{Flexible Image Transport System} {name={Flexible Image Transport System}, description={an international standard in astronomy for storing images, tables, and metadata in disk files. See the IAU FITS Standard for details}}
\newglossaryentry{Handle} {name={Handle}, description={The unique identifier assigned to a document uploaded to DocuShare}}
\newacronym{IAU} {IAU} {International Astronomical Union}
\newacronym{IVOA} {IVOA} {International Virtual Observatory Alliance}
\newglossaryentry{J2000} {name={J2000}, description={Julian Date referring to the instant of 12 noon (midday) on January 1, 2000. IAU standard equinox}}
\newacronym{JD} {JD} {Julian Date}
\newglossaryentry{Julian Date} {name={Julian Date}, description={The Julian Date (JD) of any instant is the Julian day number for the preceding noon (UTC), plus the fraction of the day elapsed since that instant. The Julian day number is a running sequence of integral days, starting at noon, since the beginning of the Julian Period; JD 0.0 corresponds to noon on 1 January 4713 BCE. Various Julian Date converters are available on the Web. For example, 18h 00m 00.0s UT on 2014-July-01 (near the start of LSST construction) corresponds to JD 2456840.25}}
\newglossaryentry{Kubernetes} {name={Kubernetes}, description={A system for automating application deployment and management using software containers (e.g. Docker); \url{https://kubernetes.io}}}
\newacronym{LDM} {LDM} {LSST Data Management (Document Handle)}
\newacronym{LPM} {LPM} {LSST Project Management (Document Handle)}
\newacronym{LSE} {LSE} {LSST Systems Engineering (Document Handle)}
\newacronym{LSST} {LSST} {Legacy Survey of Space and Time}
\newglossaryentry{LSST Project Office} {name={LSST Project Office}, description={Official name of the stand-alone AURA operating center responsible for execution of the LSST construction project under the NSF MREFC account}}
\newglossaryentry{LSST Science Pipelines} {name={LSST Science Pipelines}, description={software used to perform the LSST data reduction see \url{pipelines.lsst.io}}}
\newacronym{LSST-DA} {LSST-DA} {LSST Discovery Alliance}
\newacronym{LSSTCam} {LSSTCam} {LSST Science Camera}
\newacronym{LSSTPO} {LSSTPO} {LSST Project Office}
\newacronym{LaTeX} {LaTeX} {(Leslie) Lamport TeX (document markup language and document preparation system)}
\newacronym{MA} {MA} {Maintenance}
\newacronym{MREFC} {MREFC} {Major Research Equipment and Facility Construction}
\newglossaryentry{Major Research Equipment and Facility Construction} {name={Major Research Equipment and Facility Construction}, description={the NSF account through which large facilities construction projects such as LSST are funded}}
\newacronym{NCSA} {NCSA} {National Center for Supercomputing Applications}
\newacronym{NOIRLab} {NOIRLab} {NSF's National Optical-Infrared Astronomy Research Laboratory; \url{https://noirlab.edu}}
\newacronym{NSF} {NSF} {National Science Foundation}
\newglossaryentry{National Science Foundation} {name={National Science Foundation}, description={primary federal agency supporting research in all fields of fundamental science and engineering; NSF selects and funds projects through competitive, merit-based review}}
\newacronym{OSS} {OSS} {Observatory System Specifications; LSE-30}
\newglossaryentry{Object} {name={Object}, description={In LSST nomenclature this refers to an astronomical object, such as a star, galaxy, or other physical entity. E.g., comets, asteroids are also Objects but typically called a Moving Object or a Solar System Object (SSObject). One of the DRP data products is a table of Objects detected by LSST which can be static, or change brightness or position with time}}
\newglossaryentry{Operations} {name={Operations}, description={The 10-year period following construction and commissioning during which the Rubin Observatory conducts its survey}}
\newacronym{PB} {PB} {PetaByte}
\newacronym{PDF} {PDF} {Probability Density Function}
\newacronym{PSF} {PSF} {Point Spread Function}
\newacronym{PSTN} {PSTN} {Project Science Technical Note}
\newglossaryentry{Project Manager} {name={Project Manager}, description={The person responsible for exercising leadership and oversight over the entire Rubin project; he or she controls schedule, budget, and all contingency funds}}
\newglossaryentry{Prompt Processing} {name={Prompt Processing}, description={The data processing which occurs at the Archive Center based on the stream of images coming from the telescope. This includes both Alert Production, which scans the image stream to identify and send alerts on transient and variable sources, and Solar System Processing, which identifies and characterizes objects in our solar system. It also includes specialized rapid calibration and Commissioning processing. Prompt Processing generates the Prompt Data Products}}
\newacronym{QA} {QA} {Quality Assurance}
\newacronym{QC} {QC} {Quality Control}
\newglossaryentry{Quality Assurance} {name={Quality Assurance}, description={All activities, deliverables, services, documents, procedures or artifacts which are designed to ensure the quality of DM deliverables. This may include QC systems, in so far as they are covered in the charge described in LDM-622. Note that contrasts with the LDM-522 definition of ‚'QA' as ‚'Quality Analysis', a manual process which occurs only during commissioning and operations. See also: Quality Control}}
\newglossaryentry{Quality Control} {name={Quality Control}, description={Services and processes which are aimed at measuring and monitoring a system to verify and characterize its performance (as in LDM-522). Quality Control systems run autonomously, only notifying people when an anomaly has been detected. See also Quality Assurance}}
\newacronym{RAG} {RAG} {Retrieval-Augmented Generation}
\newacronym{REC} {REC} {Raft Electronics Crate}
\newacronym{REST} {REST} {REpresentational State Transfer}
\newacronym{RSP} {RSP} {Rubin Science Platform}
\newacronym{RTN} {RTN} {Rubin Technical Note}
\newglossaryentry{Raft} {name={Raft}, description={The sensors in the LSST camera are packaged into replaceable electronic assemblies, called rafts, consisting of 9 butted sensors (CCDs) in a 3x3 mosaic. Each raft is a replaceable unit in the LSST camera. There are 21 science rafts in the camera plus 4 additional corner rafts with specialized, non-science sensors, making for a total of 189 CCDs per focal plane image. The 21 science rafts are numbered from '0,1' through '0,3' '1,0' through '3,4' and '4,1' through '4,3' (In other words, the 25 combinations from '0,0' through '4,4' minus the four corners which are non-science.)}}
\newacronym{SLAC} {SLAC} {SLAC National Accelerator Laboratory}
\newglossaryentry{SLAC National Accelerator Laboratory} {name={SLAC National Accelerator Laboratory}, description={A national laboratory funded by the US Department of Energy (DOE); SLAC leads a consortium of DOE laboratories that has assumed responsibility for providing the LSST camera. Although the Camera project manages its own schedule and budget, including contingency, the Camera team's schedule and requirements are integrated with the larger Project.  The camera effort is accountable to the LSSTPO}}
\newacronym{SMBH} {SMBH} {Supermassive Black Hole}
\newacronym{SQR} {SQR} {SQuARE document handle}
\newacronym{SRD} {SRD} {LSST Science Requirements; LPM-17}
\newglossaryentry{Science Pipelines} {name={Science Pipelines}, description={The library of software components and the algorithms and processing pipelines assembled from them that are being developed by DM to generate science-ready data products from LSST images. The Pipelines may be executed at scale as part of LSST Prompt or Data Release processing, or pieces of them may be used in a standalone mode or executed through the Rubin Science Platform. The Science Pipelines are one component of the LSST Software Stack}}
\newglossaryentry{Simonyi Survey Telescope} {name={Simonyi Survey Telescope}, description={The telescope at the Rubin Observatory that will perform the LSST (this refers to all physical components: the mirror, the mount assembly, etc.)}}
\newglossaryentry{Software Stack} {name={Software Stack}, description={Often referred to as the LSST Stack, or just The Stack, it is the collection of software written by the LSST Data Management Team to process, generate, and serve LSST images, transient alerts, and catalogs. The Stack includes the LSST Science Pipelines, as well as packages upon which the DM software depends. It is open source and publicly available}}
\newglossaryentry{Solar System Object} {name={Solar System Object}, description={A solar system object is an astrophysical object that is identified as part of the Solar System: planets and their satellites, asteroids, comets, etc. This class of object had historically been referred to within the LSST Project as Moving Objects}}
\newglossaryentry{Solar System Processing} {name={Solar System Processing}, description={A component of the Prompt Processing system, Solar System Processing identifies new SSObjects using unassociated DIASources}}
\newglossaryentry{Source} {name={Source}, description={A single detection of an astrophysical object in an image, the characteristics for which are stored in the Source Catalog of the DRP database. The association of Sources that are non-moving lead to Objects; the association of moving Sources leads to Solar System Objects. (Note that in non-LSST usage 'source' is often used for what LSST calls an Object.)}}
\newglossaryentry{Subsystem} {name={Subsystem}, description={A set of elements comprising a system within the larger LSST system that is responsible for a key technical deliverable of the project}}
\newglossaryentry{Subsystem Manager} {name={Subsystem Manager}, description={responsible manager for an LSST subsystem; he or she exercises authority, within prescribed limits and under scrutiny of the Project Manager, over the relevant subsystem's cost, schedule, and work plans}}
\newglossaryentry{Systems Engineering} {name={Systems Engineering}, description={an interdisciplinary field of engineering that focuses on how to design and manage complex engineering systems over their life cycles. Issues such as requirements engineering, reliability, logistics, coordination of different teams, testing and evaluation, maintainability and many other disciplines necessary for successful system development, design, implementation, and ultimate decommission become more difficult when dealing with large or complex projects. Systems engineering deals with work-processes, optimization methods, and risk management tools in such projects. It overlaps technical and human-centered disciplines such as industrial engineering, control engineering, software engineering, organizational studies, and project management. Systems engineering ensures that all likely aspects of a project or system are considered, and integrated into a whole}}
\newacronym{TAP} {TAP} {Table Access Protocol (IVOA standard)}
\newacronym{TB} {TB} {TeraByte}
\newacronym{UK} {UK} {United Kingdom}
\newacronym{URL} {URL} {Universal Resource Locator}
\newacronym{US} {US} {United States}
\newacronym{USA} {USA} {United States of America}
\newacronym{USDF} {USDF} {United States Data Facility}
\newacronym{UT} {UT} {Universal Time}
\newacronym{UTC} {UTC} {Coordinated Universal Time}
\newglossaryentry{Validation} {name={Validation}, description={A process of confirming that the delivered system will provide its desired functionality; overall, a validation process includes the evaluation, integration, and test activities carried out at the system level to ensure that the final developed system satisfies the intent and performance of that system in operations}}
\newacronym{YAML} {YAML} {Yet Another Markup Language}
\newglossaryentry{airmass} {name={airmass}, description={The pathlength of light from an astrophysical source through the Earth's atmosphere. It is given approximately by sec z, where z is the angular distance from the zenith (the point directly overhead, where airmass = 1.0) to the source}}
\newglossaryentry{algorithm} {name={algorithm}, description={A computational implementation of a calculation or some method of processing}}
\newglossaryentry{astronomical object} {name={astronomical object}, description={A star, galaxy, asteroid, or other physical object of astronomical interest. Beware: in non-LSST usage, these are often known as sources}}
\newglossaryentry{calibration} {name={calibration}, description={The process of translating signals produced by a measuring instrument such as a telescope and camera into physical units such as flux, which are used for scientific analysis. Calibration removes most of the contributions to the signal from environmental and instrumental factors, such that only the astronomical component remains}}
\newglossaryentry{camera} {name={camera}, description={An imaging device mounted at a telescope focal plane, composed of optics, a shutter, a set of filters, and one or more sensors arranged in a focal plane array}}
\newglossaryentry{cloud} {name={cloud}, description={A visible mass of condensed water vapor floating in the atmosphere, typically high above the ground or in interstellar space acting as the birthplace for stars.  Also a way of computing (on other peoples computers leveraging their services and availability)}}
\newglossaryentry{configuration} {name={configuration}, description={A task-specific set of configuration parameters, also called a `config'. The config is read-only; once a task is constructed, the same configuration will be used to process all data. This makes the data processing more predictable: it does not depend on the order in which items of data are processed. This is distinct from arguments or options, which are allowed to vary from one task invocation to the next}}
\newglossaryentry{cycle} {name={cycle}, description={The time period over which detailed, short-term plans are defined and executed. Normally, cycles run for six months, and culminate in a new release of the LSST Software Stack, however this need not always be the case}}
\newglossaryentry{epoch} {name={epoch}, description={Sky coordinate reference frame, e.g., J2000. Alternatively refers to a single observation (usually photometric, can be multi-band) of a variable source}}
\newglossaryentry{flux} {name={flux}, description={Shorthand for radiative flux, it is a measure of the transport of radiant energy per unit area per unit time. In astronomy this is usually expressed in cgs units: erg/cm\$\^2\$/s}}
\newglossaryentry{metadata} {name={metadata}, description={General term for data about data, e.g., attributes of astronomical objects (e.g. images, sources, astroObjects, etc.) that are characteristics of the objects themselves, and facilitate the organization, preservation, and query of data sets (e.g., a FITS header contains metadata)}}
\newglossaryentry{middleware} {name={middleware}, description={Software that acts as a bridge between other systems or software usually a database or network. In the Data Management System this refers to Butler for data access and Workflow management for distributed processing}}
\newglossaryentry{monitoring} {name={monitoring}, description={In DM QA, this refers to the process of collecting, storing, aggregating and visualizing metrics}}
\newglossaryentry{patch} {name={patch}, description={An quadrilateral sub-region of a sky tract, with a size in pixels chosen to fit easily into memory on desktop computers}}
\newglossaryentry{pipeline} {name={pipeline}, description={A configured sequence of software tasks (Stages) to process data and generate data products. Example: Association Pipeline}}
\newglossaryentry{provenance} {name={provenance}, description={Information about how LSST images, Sources, and Objects were created (e.g., versions of pipelines, algorithmic components, or templates) and how to recreate them}}
\newglossaryentry{schema} {name={schema}, description={The definition of the metadata and linkages between datasets and metadata entities in a collection of data or archive}}
\newglossaryentry{shape} {name={shape}, description={In reference to a Source or Object, the shape is a functional characterization of its spatial intensity distribution, and the integral of the shape is the flux. Shape characterizations are a data product in the DIASource, DIAObject, Source, and Object catalogs}}
\newglossaryentry{sky map} {name={sky map}, description={A sky tessellation for LSST. The Stack includes software to define a geometric mapping from the representation of World Coordinates in input images to the LSST sky map. This tessellation is comprised of individual tracts which are, in turn, comprised of patches}}
\newglossaryentry{software} {name={software}, description={The programs and other operating information used by a computer}}
\newglossaryentry{stack} {name={stack}, description={a grouping, usually in layers (hence stack), of software packages and services to achieve a common goal. Often providing a higher level set of end user oriented services and tools}}
\newglossaryentry{tract} {name={tract}, description={A portion of sky, a spherical convex polygon, within the LSST all-sky tessellation (sky map). Each tract is subdivided into sky patches}}
\newglossaryentry{transient} {name={transient}, description={A transient source is one that has been detected on a difference image, but has not been associated with either an astronomical object or a solar system body}}

%% file: authors.tex

\author[a]{Leanne~P.~Guy}
\author[b]{Connor~Yablonski}
\author[c]{Aaron~M.~Meisner}
\author[d]{Guillem~Megias~Homar}
\author[e]{Merlin~Fisher-Levine}
\author[f]{Eman~E.~Ali}
\author[g]{Tiger~J.~Hu}
\author[h,i,j]{Christopher~W.~Stubbs}
\affil[a]{NSF-DOE Vera C.\ Rubin Observatory / NSF NOIRLab, Casilla 603, La Serena, Chile}
\affil[b]{University of Washington, Dept.\ of Astronomy, Box 351580, Seattle, WA 98195, USA}
\affil[c]{NSF-DOE Vera C.\ Rubin Observatory / NSF NOIRLab, 950 N.\ Cherry Ave., Tucson, AZ  85719, USA}
\affil[d]{Division of Physics, Mathematics and Astronomy, California Institute of Technology, Pasadena, CA 91125, USA}
\affil[e]{D4D CONSULTING LTD., Suite 1 Second Floor, Everdene House, Deansleigh Road, Bournemouth, UK BH7 7DU}
\affil[f]{Swinburne University of Technology, Melbourne, Victoria, Australia}
\affil[g]{Australian Astronomical Optics, Macquarie University, North Ryde, NSW, Australia}
\affil[h]{Department of Astronomy, Center for Astrophysics, Harvard University, 60 Garden St., Cambridge, MA 02138, USA}
\affil[i]{Center for Astrophysics, Harvard \& Smithsonian, 60 Garden Street, Cambridge, MA 02138}
\affil[j]{Department of Physics, Harvard University, 17 Oxford St., Cambridge MA 02138, USA}

%% file: abstract.tex
\begin{abstract}
The NSF-DOE Vera C. Rubin Observatory will generate petabytes of data through the Legacy Survey of Space and Time (LSST) over the next decade, enabling discoveries across a broad range of astrophysical fields. 
Alongside these data products, Rubin maintains a large but heterogeneous collection of supporting documentation, including operational guides, technical notes, and scientific papers. 
Because this material is distributed across multiple platforms and formats, staff and scientists often struggle to efficiently locate accurate, up-to-date information. 
Many resources also reside on internal systems, limiting the ability of general-purpose language models to provide reliable answers to Rubin-specific questions. 
To address these challenges, we explore the use of Retrieval-Augmented Generation (RAG) to improve information discovery. 
We present a prototype RAG-based virtual assistant that delivers context-aware, factual, conversational access to Rubin’s vast and heterogenous documentation ecosystem. 
The system integrates material from multiple sources and enables semantic search through a conversational interface, using Weaviate for embeddings, LangChain for query orchestration, and an OpenAI GPT model as the LLM backend. 
By grounding responses in domain-specific knowledge, the assistant reduces hallucinations, improves accuracy, and demonstrates the potential of RAG to enhance access to distributed knowledge, streamline workflows, and support effective use of LSST data products.
\end{abstract}

%% file: body.tex
\section{Introduction}
\label{sec:intro}
The NSF-DOE Vera C. Rubin Observatory\cite{2019ApJ...873..111I}, located at an elevation of 2647\,m on Cerro Pach\'{o}n in Chile, is a large-aperture, wide-field optical observatory whose prime mission is to conduct the 10-year \gls{LSST}, beginning in 2026.
The \gls{LSST} will repeatedly image approximately 18,000 square degrees of sky every three nights using a 3.2-gigapixel camera mounted on the 8.4-meter \gls{Simonyi Survey Telescope}.
Each night, Rubin will generate 15--20\,TB of raw data from which it will issue approximately 10 million \gls{transient} alerts.
Over its lifetime, the \gls{LSST} is expected to catalog approximately 20 billion galaxies and 17 billion stars, producing a final catalog of approximately 15\,PB and roughly 500\,PB of image products across several major data releases, each accompanied by updated documentation, data product definitions, and software.

This scientific output is accompanied by an extensive body of supporting documentation, including technical notes, design documents, operational guides, data product definitions, software documentation, and community-facing tutorials, distributed across a wide variety of platforms: Confluence wikis, Jira tickets, GitHub repositories, the Rubin Community Forum\footnote{\url{https://community.lsst.org}}, Slack discussions, \gls{DocuShare} archives, and numerous PDF reports.
This fragmentation creates a significant challenge for both Rubin Observatory staff and the broader science community when attempting to locate accurate, up-to-date information.
The challenge is compounded by the scale of the Rubin science community itself: approximately 10,000 scientists across 28 countries hold Rubin data rights, a regime in which traditional helpdesk-based user support models do not scale\cite{RTN-097}.
Rubin's approach to community science instead focuses on fostering a vibrant community supported by infrastructure such as the Discourse-based Rubin Community Forum, to crowdsource solutions and build a deep reservoir of shared expertise\cite{RTN-121}.

Large language models (LLMs) offer a promising approach to this information-discovery challenge.
LLMs are language models trained on large datasets capable of performing a variety of text- and language-related tasks.
Those that generate novel text in response to a user prompt are called \emph{generative} LLMs; those that encode text into a semantic vector space are called \emph{embedding} models.
Their ability to understand and generate natural language makes generative LLMs well suited to serving as an intelligent interface to large, heterogeneous documentation ecosystems, allowing users to find information through conversation without needing to know where each piece of information resides.
However, both classes of model face fundamental limitations when applied to domain-specific questions about Rubin Observatory.
The knowledge encoded in model weights during training is \emph{static}: it cannot be updated as new information becomes available, and models may already be outdated by the time they are released.
Because this training draws on broad general corpora, models often lack the \emph{domain-specific} knowledge needed to answer questions about Rubin's data products, the \gls{RSP}, or the \gls{LSST} science pipelines~\cite{PSTN-019}, and generative models may produce plausible but factually incorrect responses, a phenomenon known as \emph{hallucination}.
Furthermore, much of the information critical to Rubin operations and science resides on internal or proprietary systems, making it entirely inaccessible to general-purpose models.

The domain itself presents additional retrieval challenges that affect the full information-discovery pipeline for both observatory staff and the science community.
Astrophysical vocabulary is dense and highly specialized, and the relevance of a concept often depends strongly on context.
Terms such as ``AGN feedback,'' ``SMBH accretion,'' ``halo quenching,'' and ``Type~Ia progenitor channels'' refer to distinct physical processes, yet in the context of a query about how compact objects influence galaxy evolution, all may be highly relevant despite sharing little direct keyword overlap.
Mathematical symbols and equations carry semantic weight that general-purpose retrieval systems may not fully capture.
In the Rubin context, these challenges are compounded by observatory-specific terminology in which common English words carry precise technical meanings (e.g., ``Butler'' as a data access framework, ``tract'' and ``patch'' as sky tiling units) and by a knowledge base spanning staff operations, engineering, and science across dozens of platforms.

\gls{RAG} directly addresses both the staleness and accessibility limitations of standalone LLMs by pairing a generative model with an external, searchable data source from which relevant passages are fetched at query time, grounding responses in specific, current information rather than the model's internal approximations.
\gls{RAG} was introduced by Lewis et al.~\cite{NEURIPS2020_6b493230} to formalize this approach: their key insight was that LLMs possess implicit \emph{parametric} memory encoded in model weights, but that combining it with explicit \emph{non-parametric} memory in the form of a searchable document index significantly improves factual accuracy on knowledge-intensive tasks.
RAG was selected for this application over model fine-tuning because the core challenge is not how the model responds but what information it can access, and 
because Rubin's knowledge base must evolve across data releases without expensive retraining~\cite{RAGvsFT}.

In this paper we present a \gls{RAG}-based virtual assistant designed to ground LLM responses in Rubin Observatory's knowledge bases, providing accurate, current, and contextually relevant answers while reducing hallucinations.
The system targets two complementary user populations: Rubin Observatory staff who need to rapidly locate information across the project's myriad platforms and formats, and the global Rubin science community who need guidance on exploiting Rubin data products and services.
We describe the motivation, architecture, and implementation of a prototype system, the lessons learned during its development, and future directions.

\section{Motivation and Use Cases}
\label{sec:motivation}

\subsection{Supporting daily Rubin operations}
Rubin Observatory staff rely daily on a broad set of private collaboration resources, including Confluence wiki pages, Jira tickets, Slack discussions, Community Forum threads, proprietary documents, engineering drawings, and a long history of design documents in the internal \gls{DocuShare} archive.
A \gls{RAG}-grounded virtual assistant that makes this distributed and rapidly evolving body of knowledge searchable through a single conversational interface would significantly reduce the time staff spend locating information and help prevent operational decisions from being made on the basis of outdated or superseded documents.
Making this distributed and rapidly evolving body of knowledge searchable through a single conversational interface would significantly reduce the time staff spend locating information and help prevent operational decisions being made on the basis of outdated or superseded documents.
A particular strength of \gls{RAG} in this context is its ability to synthesize information across platforms.
For example, a question such as \emph{``what observing strategy was adopted for the Rubin commissioning science validation surveys?"} might require combining the technote reporting on the \gls{LSSTCam} observing campaign strategy,  Jira tickets recording the actual commands executed, and Slack threads where details were discussed.
No single platform contains the complete answer; a \gls{RAG} system searching across all three can assemble it.
Similarly, ingesting the \gls{Simonyi Survey Telescope} troubleshooting guide from Confluence would allow night-time observers to query telescope faults in natural language during time-critical operations.
Key reference documents such as the \gls{LSST} Science Book and the final telescope design document can likewise be made conversationally accessible, with answers traceable to their source.
The virtual assistant would also support onboarding of new Observatory staff, who must rapidly absorb institutional knowledge scattered across over 20 years of archived discussions, technical notes, and design decisions.
Staff queries also present intrinsic retrieval challenges: technical vocabulary is dense and context-dependent, and the same subsystem, component, or failure mode is frequently referred to by different names across platforms. Jira tickets, Confluence pages, and Slack threads each develop their own shorthand for the same concepts, requiring a retrieval system that matches on meaning rather than exact terms.

\subsection{Supporting the science community}
The Rubin \gls{RAG} virtual assistant is also designed to support the global science community in exploiting Rubin data products.
The Rubin Early Science Program\cite{RTN-011} will release science-grade commissioning and early pre-survey data to support  community preparations for Rubin operations and enable high-impact science as early as possible.
The first full \gls{LSST} Data Release, \gls{DR1}, is expected approximately two years after the start of survey operations.
In the intervening period, a substantial volume of data will be made available through transient alerts, nightly processed images, and catalogs of single-epoch detections, as well as through the Data Preview program (\gls{DP1}\cite{RTN-095}, \gls{DP2}) in the period leading up to \gls{DR1}.
Each of these releases represents an opportunity for early scientific discovery with Rubin data.
During this same period, the community will also be familiarizing itself with Rubin's data products, services, access tools, and the algorithms used to produce the data products.
A virtual assistant grounded in authoritative, up-to-date Rubin documentation can lower the barrier to entry, help users navigate unfamiliar systems, and accelerate the path from data access to scientific results.

The eventual goal is for the Rubin virtual assistant to answer technical questions and provide at least beginner-level coding assistance for working with the \gls{LSST} Science Pipelines.
A key challenge for users will be navigating the Rubin data model\cite{LSE-163}, which comprises hundreds of columns across multiple catalog tables with Rubin-specific naming conventions, and constructing correct and efficient ADQL/TAP queries.
For example, a user asking \emph{``how do I select clean galaxy samples from the Object table?''} needs to know which flag columns to filter on to exclude objects with unreliable shape or light profile modeling, and how to apply signal-to-noise cuts to retain only well-detected galaxies. These column names are specific to Rubin and may change between data releases; a general-purpose LLM may hallucinate plausible but incorrect names, while a \gls{RAG} system grounded in the current schema helps mitigate this risk.
Similarly, a question such as \emph{``how do I retrieve the coadded images for the ECDFS field from DP1?"} requires knowing the correct Butler\cite{2022SPIE12189E..11J} collection name, the filter designations available in DP1, and the current Butler API syntax, all details that may change between releases and are absent from general LLM training data.
A \gls{RAG} system grounded in the schema documentation, data product definitions, and worked examples from tutorials and Community Forum posts can provide reliable, validated guidance that a general-purpose LLM cannot.
The Community Forum is a particularly valuable knowledge source: a user encountering an error running a tutorial notebook may find that the identical problem was already resolved on the Forum; the virtual assistant can surface that answer directly rather than requiring the user to search the Forum manually.

\subsection{Documentation validation}
A \gls{RAG} virtual assistant also serves as a tool for validating Rubin's user-facing documentation.
By creating a standard suite of prompts and systematically querying the virtual assistant, the team can identify gaps, inconsistencies, or ambiguities in existing documentation.
If the virtual assistant cannot answer a question that should be answerable, this signals a documentation gap.
If it produces an incorrect answer, this may indicate ambiguous or contradictory source material.
For example, testing whether the virtual assistant correctly reports which filters are available in \gls{DP1}, a subset of the full \textit{ugrizy} set, can reveal whether the documentation clearly distinguishes \gls{DP1}-specific content from general \gls{LSST} descriptions.
The approach can also detect whether tutorial notebooks reference deprecated API calls that no longer work against the current software stack, or whether documented Butler collection names match the collections actually deployed on the RSP.
Rubin's Community Science Team has already explored using \href{https://vale.sh}{Vale} for automated style verification of user-facing documentation; the virtual assistant provides a complementary, content-level validation approach.
%
\section{Architecture}
\label{sec:architecture}
 
\begin{figure}[t]
  \centering
  \includegraphics[width=0.85\textwidth]{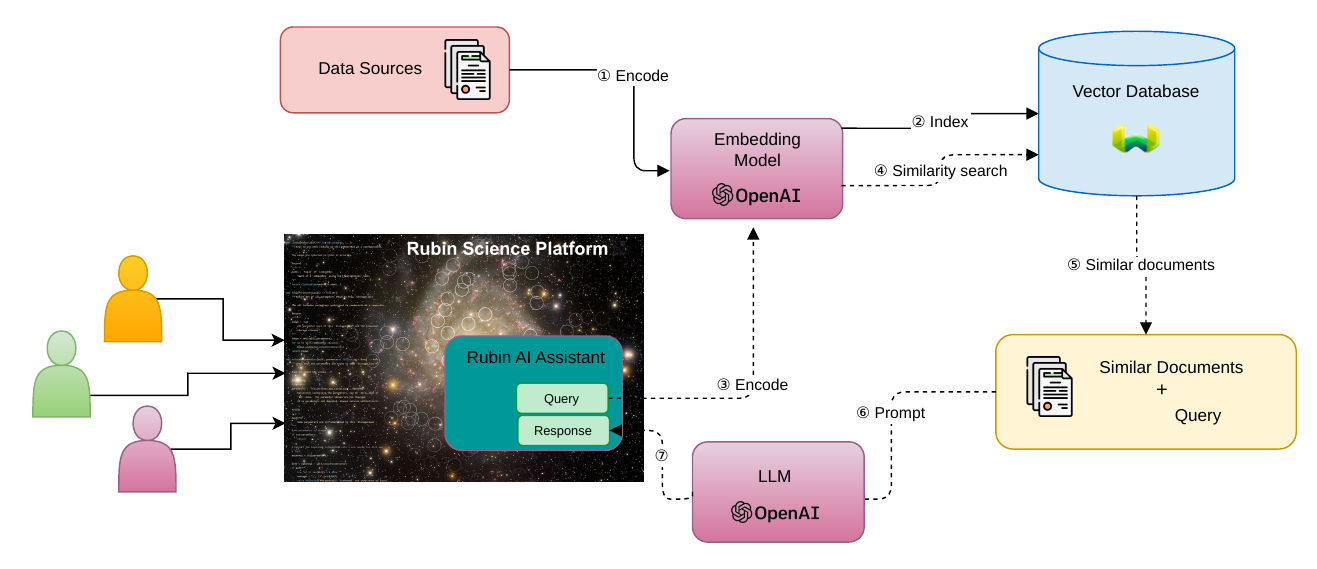}
  \caption{Architecture of the Rubin AI virtual assistant showing the flow from document ingestion through embedding, vector storage, retrieval, and response generation. Documents from the Rubin knowledge base are loaded, chunked, and embedded using OpenAI's \texttt{text-embedding-3-large} model. Embeddings and text chunks are stored in the Weaviate\cite{weaviate} vector database.
  At query time, a user query submitted via the Streamlit frontend is embedded using the same model; the top-$k$ most similar chunks are retrieved and injected into the LLM prompt to produce a grounded response. The system is deployed on the \gls{RSP} through the Phalanx infrastructure.}
  \label{fig:architecture}
\end{figure}

We adopt the modular variant of the RAG architecture\cite{Gao2024}, in which the retrieval, augmentation, and generation components are independently built, tested, and swappable.
This modularity has been essential during development, allowing us to experiment with different embedding models, chunking strategies, and LLM backends without redesigning the full system.
Figure~\ref{fig:architecture} shows the overall system architecture.
 
\subsection{Overview: the RAG framework}
\label{sec:rag_framework}

The three words in Retrieval-Augmented Generation describe the three stages of the system:

\begin{itemize}
\item\textbf{Retrieval:} Given a user query, the system searches an external knowledge base to find the most relevant document fragments. The user's query is converted into a vector embedding and compared against stored document embeddings using cosine similarity to identify the top-$k$ most semantically relevant chunks. 
\item\textbf{Augmentation:} The retrieved chunks are assembled into a structured context block and combined with the original user query to form an augmented prompt. A system prompt instructs the LLM to base its answer on the provided context and to indicate when the available documentation does not contain sufficient information, rather than generating a speculative response. This augmentation step is what distinguishes \gls{RAG} from a standard LLM interaction: the model receives domain-specific context alongside the question.
\item\textbf{Generation:} The augmented prompt is passed to a generative LLM, which produces a natural language response conditioned on both the user's question and the retrieved context. Because the response is grounded in specific, retrieved documents rather than the model's compressed parametric knowledge alone, the output is more factually accurate and traceable to its sources.
\end{itemize}

The following subsections describe the specific components selected for each stage, together with the frontend, deployment infrastructure, and platform integration that connect them into a working service.
 
\subsection{Retrieval: Weaviate vector database}
\label{sec:weaviate}
 
The retrieval stage depends on an efficient, searchable store of document embeddings.
We use \href{https://weaviate.io}{Weaviate}~\cite{weaviate} as the vector database.
Weaviate is built around vector embeddings as the primary data type, with storage layout and query planning designed for approximate nearest neighbor search over high-dimensional spaces; embedding models are integrated at the data layer, so vectors are generated and indexed at ingest rather than as a post-processing step.
Weaviate organizes data in ``collections'' that store text chunks, associated metadata (source URL, document identifier, page number, creation date), and their vector embeddings together.
This co-location of vectors, text, and metadata allows the retrieval component to efficiently compare the vector embedding of a user query against the stored embeddings using cosine similarity, and to return both the matching text and its provenance.
Weaviate was selected for its native integration with LangChain\cite{langchain}, its support for multiple vectorization backends, its native hybrid search capability combining dense vector and sparse keyword search, and its suitability for deployment on \gls{Kubernetes}.
 
\subsection{Retrieval and generation: embedding and generative models}
\label{sec:llm}
 
The system uses two separate OpenAI models serving distinct roles in the retrieval and generation stages.
The \texttt{text-embedding-3-large} model is used as the \emph{vectorizer} in the retrieval stage, converting both document chunks and user queries into high-dimensional vector representations that capture semantic meaning.
The same embedding model is used at ingestion time and at query time, ensuring that documents and queries occupy the same vector space and that similarity comparisons are meaningful.
The \texttt{gpt-3.5-turbo} model serves as the \emph{generative LLM} in the generation stage, producing natural language responses based on the augmented prompt containing the user's query and the retrieved context.
Both models are accessed via the OpenAI API through keys provided to the Weaviate client and LangChain.
 
 
While we use OpenAI for both components during development, the modular architecture allows alternative LLMs to be substituted.
We plan to benchmark alternative LLMs including more recent OpenAI models, Anthropic's Claude, and open-source alternatives to optimize the trade-off between response quality, latency, and cost.
 
\subsection{Augmentation and orchestration: LangChain}
\label{sec:langchain}
 
\href{https://www.langchain.com}{LangChain} is an open-source Python library that provides the integration layer connecting all three \gls{RAG} stages.
During ingestion, LangChain manages the document loading, chunking, and embedding pipeline that populates the Weaviate vector store (Section~\ref{sec:ingestion}).
At query time, LangChain orchestrates the full \gls{RAG} cycle: embedding the user query, retrieving relevant chunks from Weaviate, constructing the augmented prompt with retrieved context, and routing the augmented prompt to the generative LLM.
Its modular design allows individual components, including loaders, splitters, embedding models, vector stores, and LLMs, to be swapped independently, which has been essential for the iterative experimentation and optimization described in Section~\ref{sec:future}.
 
\subsection{Frontend: Streamlit}
\label{sec:frontend}
 
\begin{figure}[ht]
  \centering
  \includegraphics[width=0.85\textwidth]{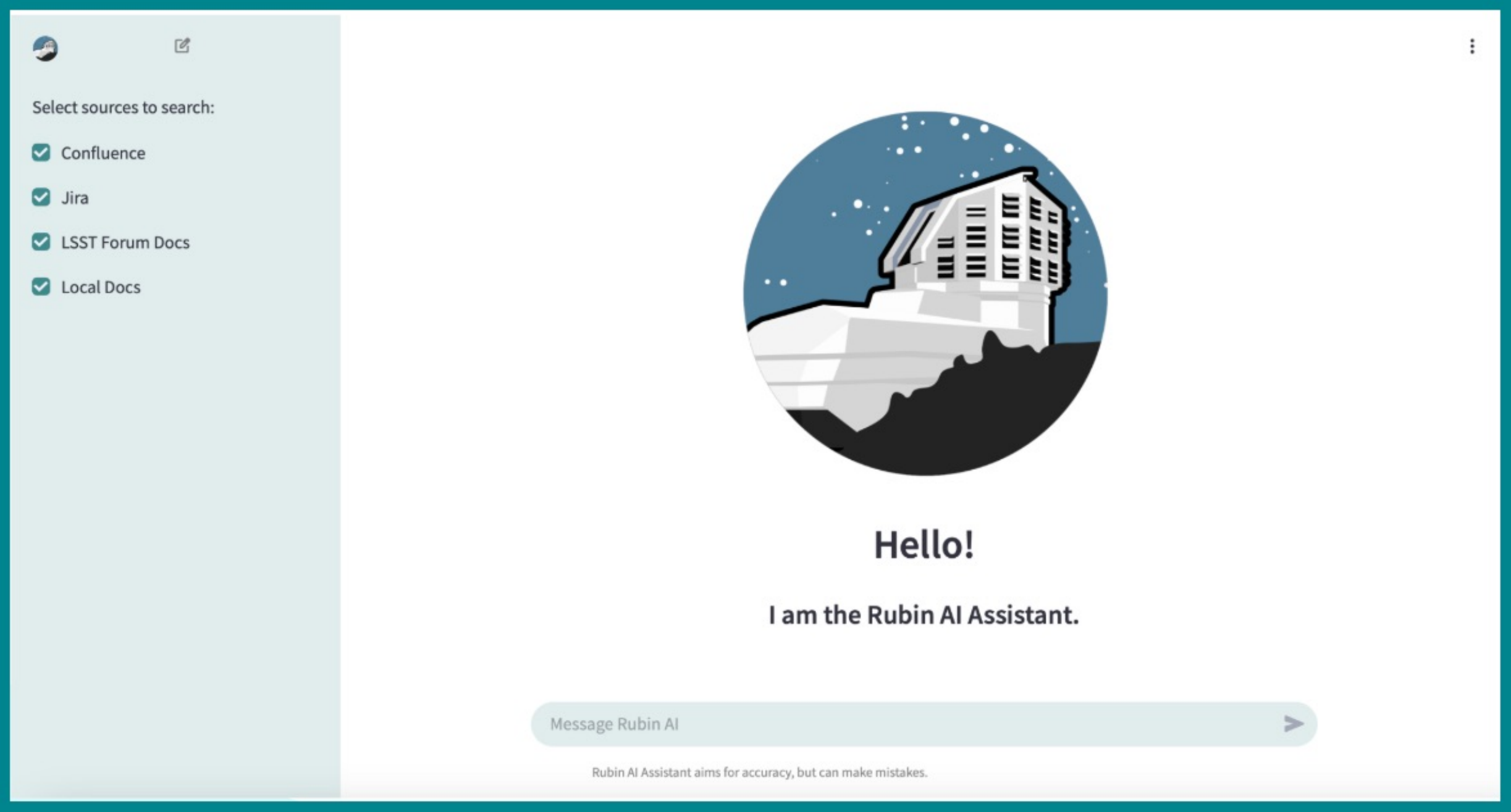}
  \caption{The Rubin AI Assistant conversational interface, implemented in Streamlit and deployed on the \gls{RSP}. The left sidebar shows the source-filter panel (Confluence, Jira, \gls{LSST} Forum Docs, Local Docs), allowing users to target specific document collections. Users enter natural-language questions in the message bar; the system retrieves relevant document chunks and returns a grounded response with source citations.}
  \label{fig:rubin_rag_ui}
\end{figure}

The user-facing interface is implemented as a \href{https://streamlit.io}{Streamlit} application providing a conversational chat interface.
The interface includes selectable source filters, e.g Confluence, Jira, \gls{LSST} forum,  that allow users to control which document collections are searched for a given query.
The application is lightweight, requiring relatively little code while providing an intuitive chat paradigm.
Figure~\ref{fig:rubin_rag_ui} shows the interface as deployed on the \gls{RSP}.

\subsection{Deployment infrastructure}
\label{sec:deployment}
 
The virtual assistant is deployed using Rubin Observatory's standard deployment model, Phalanx \cite{SQR-056}. 
Phalanx is built on a layered infrastructure stack: \href{https://kubernetes.io}{Kubernetes} for container orchestration, \href{https://argo-cd.readthedocs.io}{Argo~CD} for GitOps-based continuous delivery, \href{https://helm.sh}{Helm} for packaging \gls{Kubernetes} applications, and the \href{https://safir.lsst.io}{Safir} framework for Rubin-specific service development.
The virtual assistant is containerized via a Dockerfile in the repository and deployed through the same GitOps workflows used for all other \gls{RSP} services.
 
The Phalanx-based deployment model means that the virtual assistant's configuration is managed as code in the \href{https://github.com/lsst-sqre/phalanx}{Phalanx repository}, enabling reproducible deployments, version-controlled updates, and rollback capability.
  
\subsection{Integration with the Rubin Science Platform}
\label{sec:rsp_integration}
 
From the outset, the RAG virtual assistant was designed to be deployed as a service within the \gls{RSP}, following the same model used for all other RSP services, at both the user-facing cloud RSP\footnote{\url{https://data.lsst.cloud}} and the on-premises United States Data Facility (USDF) at SLAC National Accelerator Laboratory\cite{2024ASPC..535..227O}.

The \gls{LSST} will accumulate approximately 50\,PB of data products by the end of year~1, growing to roughly 500\,PB by year~10; far too large for users to download and analyse locally.
The \gls{RSP} is Rubin's answer to this challenge, providing cloud-based access to \gls{LSST} data products and services through three interfaces: the Portal (browser-based data discovery), Notebooks (Jupyter-based analysis), and Web APIs (programmatic access via Virtual Observatory interfaces).
Deploying the virtual assistant within this same infrastructure provides several important advantages:

\begin{itemize}
\item\textbf{Co-location with the data}: Users interact with the virtual assistant in the same environment where they access and analyze \gls{LSST} data products, enabling a tightly integrated workflow in which documentation assistance and data analysis are available side by side.
\item\textbf{Shared authentication and access control}: The virtual assistant inherits the RSP's identity management and access control infrastructure, ensuring that private or restricted resources are accessible only to authorized users without requiring separate login.
\item\textbf{Operational monitoring}: The virtual assistant benefits from Rubin's existing service monitoring and alerting infrastructure, reducing the operational burden of running an additional service.
\item\textbf{Dual-environment coverage}: Deployment at both the cloud RSP and the \gls{USDF} ensures that the virtual assistant is accessible to users regardless of which environment they use for data analysis.
\end{itemize}

\subsection{Codebase}
\label{sec:codebase}
 
The codebase is developed in Python and maintained as open-source software in the \href{https://github.com/lsst-dm/rubin_rag}{\texttt{lsst-dm/rubin\_rag}} GitHub repository.
The repository contains the Streamlit application, the Weaviate connection and query layer, the data ingestion code, and Jupyter notebooks used for prototyping and testing.
The deployment configuration is managed through a corresponding branch in the Phalanx repository.
The project is released under the MIT license.

\section{Data Ingestion Pipeline}
\label{sec:ingestion}

The data ingestion pipeline transforms raw documents from diverse sources into vector embeddings stored in Weaviate, ready for retrieval.
The pipeline consists of three main steps: loading and cleaning, chunking, and embedding.

\subsection{Data Sources}

The Rubin \gls{RAG} system ingests documents from the following sources:
\begin{itemize}
  \item \textbf{Rubin technical documents}: Design documents, studies, reports, requirements specifications, and technical notes authored in a variety of formats including \LaTeX{}, reStructuredText, Markdown, and occasionally Word. Where the original source is available it is ingested directly; a rendered PDF is used only when no source format can be obtained.
  \item \textbf{GitHub repositories}: Including the \gls{LSST} codebase and its documentation, Data Release tutorials and documentation;
 \item \textbf{Jira tickets:} Technical project management and issue-tracking records containing design discussions, implementation decisions, and bug reports. Administrative and access-restricted tickets are excluded.  \item \textbf{Rubin Community Forum}: Discourse-based forum threads, including resolved support questions that represent practical knowledge and experience;
 \item \textbf{Confluence:} Selected internal wiki pages containing meeting notes, design discussions, and operational procedures. Access-restricted and private spaces are excluded from ingestion.
  \item \textbf{Slack channels:} Selected technical discussion channels containing decisions, troubleshooting conversations, and domain expertise relevant to operations and data processing. Private and administrative channels are excluded; and
 \item \textbf{DocuShare:} Legacy document management archive containing historical design documents, engineering drawings, requirements documents, and early project records, often in non-standard formats. Ingesting this material presents particular challenges due to format diversity but provides access to institutional knowledge not available elsewhere.
\end{itemize}

\subsection{Loading and Cleaning}

We use LangChain document loaders to extract text from files or to parse data returned by APIs.
For PDFs, we use the \texttt{PyMuPDFLoader}, which parses text and tables effectively but does not recover non-text information embedded in images.
Each loader outputs a list of LangChain \texttt{Document} objects containing the extracted text and associated metadata (source file, creation date, page number).
For GitHub repositories we use the \texttt{GithubFileLoader}, which retrieves files directly from a specified repository and branch via the GitHub API; for Jira we query the Rubin Jira REST API directly to retrieve ticket content and metadata.
In both cases the result is a list of LangChain \texttt{Document} objects containing extracted text and metadata, ready for downstream chunking and embedding.

Several areas for improvement have been identified.
The current pipeline loses information encoded in figures and diagrams, which are prevalent in Rubin documentation; vision-language models could generate text descriptions of these for embedding, making visual content searchable.
For scanned documents and image-based PDFs, particularly prevalent in the \gls{DocuShare} legacy archive where non-PDF source formats are often unavailable, we are evaluating OCR and multimodal extraction techniques to recover text that standard PDF parsers cannot access.
Where source formats are available, we ingest them directly: LaTeX technotes are parsed from their LaTeX source rather than from rendered PDFs, and Markdown and reStructuredText documents handle tabular content reliably without the extraction issues that affect PDF tables.
For conversational sources such as the Community Forum, Jira, and Slack, raw ingestion produces noisy content mixing resolved answers with tangential discussion; extracting clean, structured records, such as resolved question-answer pairs from forum threads or decision records from Jira tickets, would substantially improve retrieval quality.
For Jira in particular, distinguishing currently relevant tickets from obsolete ones is an open challenge; approaches under consideration include filtering by ticket status and recency, weighting by activity level, and applying time-decay factors during retrieval so that older, unresolved tickets are ranked lower than recent, resolved ones.
Finally, as the corpus grows, deduplication across sources becomes important: the same information often appears in a Jira ticket, a Confluence page, and a technote, and retrieving contradictory versions of the same content degrades response quality

\subsection{Chunking}
\label{sec:chunking}

Loaded documents are split into chunks using LangChain's \texttt{RecursiveCharacterTextSplitter}, which divides text by character count while preferring to split at paragraph breaks, then line breaks, then sentence boundaries.
The current prototype uses a chunk size of 1000 characters with an overlap of 50 characters between adjacent chunks.

Chunk size represents a critical trade-off. 
Chunks that are \emph{too small} may not contain enough context for the LLM to generate a meaningful answer, even when the correct passage is successfully retrieved. 
Conversely, chunks that are \emph{too large} can cause the query embedding and chunk embedding to diverge semantically, reducing retrieval precision and increasing the likelihood of retrieving content from the wrong part of the document.

We plan to evaluate several alternatives to the current character-based chunking strategy, including hierarchical retrieval (also known as small-to-big or parent-child retrieval),\cite{LlamaIndex2024} in which large semantically coherent document chunks are split into smaller child chunks linked to their parent via unique identifiers.
At query time, the smaller child chunks are matched against the query for precision, and the corresponding parent chunks are passed to the LLM for generation, combining the accuracy of fine-grained retrieval with the richer context needed for well-grounded answers.
This decouples the granularity used for retrieval from the context provided to the generator, addressing both failure modes simultaneously.
We also plan to evaluate semantic chunking,\cite{Kamradt2024} which detects topic boundaries by measuring embedding similarity between adjacent sentences rather than splitting at fixed character counts, an approach that has been shown to improve retrieval recall over fixed-size methods\cite{Smith2024}.
Document-structure-aware and code-aware chunking, best practices now implemented in LangChain's specialized splitters,\cite{LangChain2023} use the inherent structure of source material to define chunk boundaries: section headers in technotes, function definitions in API documentation, and cell boundaries in Jupyter notebooks. 
Cell-level chunking of notebooks is particularly important, as tutorial notebooks are among the most valuable sources for community users and each cell typically represents a self-contained, executable step in a scientific workflow that naive character-based splitting would break into unusable fragments.

\subsection{Embedding}
After chunking, documents are embedded using the OpenAI \texttt{text-embedding-3-large} model via LangChain's \texttt{WeaviateVectorStore}, which creates and populates the corresponding Weaviate collections. 
Each collection stores the chunk text, associated metadata, and vector embedding together, enabling efficient similarity search at query time.
The \texttt{text-embedding-3-large} model is well suited to the semantic retrieval requirements of this application, particularly for heterogeneous scientific and technical text \cite{openai2024embeddings}. 
The same embedding model is used both during document ingestion and at query time, ensuring that documents and user queries are represented within a shared semantic vector space.

We are also evaluating metadata-enriched embedding, in which document context such as the source title, section header, and document type is prepended to each chunk prior to embedding. 
This allows the resulting vector representation to incorporate not only the chunk content itself, but also aspects of its provenance and document context, which may improve retrieval precision when source or section information is relevant to a query.

In the near term, we do not anticipate changing the embedding model unless adoption of an alternative LLM backend necessitates it, as discussed in Section~\ref{sec:nearterm}.

\section{Query Processing and Response Generation}
\label{sec:query}

When a user submits a query through the Streamlit interface, the system processes it through the following steps:

\begin{enumerate}
  \item \textbf{Query embedding}: The user's query is embedded using the same \texttt{text-embedding-3-large} model used during ingestion, producing a query vector in the same semantic space as the stored document embeddings.
  \item \textbf{Retrieval}: The query vector is compared against all stored chunk embeddings in Weaviate using cosine similarity search. The top-$k$ most similar chunks are retrieved, along with their source metadata (document title, URL, page number). Users can filter which source collections are searched, allowing targeted retrieval when the relevant domain is known.
  \item \textbf{Prompt augmentation}: The retrieved chunks are injected into the LLM's prompt alongside the original user query, forming an augmented prompt that grounds the LLM's response in specific retrieved context.
  \item \textbf{Response generation}: The \texttt{gpt-3.5-turbo} model generates a natural language response conditioned on both the user's query and the retrieved context. The response is streamed back to the user through the Streamlit interface.
  \item \textbf{Source citation}: The interface displays the source documents from which context was retrieved, enabling the user to verify the information and consult the primary sources directly.
\end{enumerate}

The current retrieval step uses pure cosine similarity search, which is effective for semantic matching but can miss queries containing exact identifiers such as column names, API calls, or error messages that do not paraphrase well.
We plan to investigate replacing this with hybrid retrieval, combining dense vector similarity with sparse BM25 keyword search.
BM25 is a classical keyword-based retrieval algorithm that ranks documents by how well they match query terms, and remains one of the most widely used sparse retrieval methods in information retrieval and search engines\cite{Robertson2009}.
In modern RAG systems, combining BM25 with dense vector search is a well-established pattern precisely because the two methods are complementary: vector search captures semantic similarity while BM25 captures exact term matches, so a chunk can surface for either reason or both\cite{Luan2021}.
Weaviate supports hybrid search natively, and this change requires no alteration to the ingestion pipeline nor embedding model.

\section{Evaluation and Current Status}
\label{sec:evaluation}
 
Evaluating \gls{RAG} systems is non-trivial: standard LLM benchmarks do not measure retrieval quality, and the correctness of a generated response depends on both the relevance of the retrieved chunks and the quality of generation conditioned on that context.
 
\subsection{Validation query set}
\label{sec:validation_framework}

We have developed a validation query set categorized into three types:

\begin{enumerate}
\item Queries answerable from general LLM knowledge without any domain-specific retrieval, to verify baseline behavior, for example: \emph{``What is a color-magnitude diagram?''}.
\item Queries that require Rubin-specific documentation to answer correctly, to verify that retrieval is effective and the system outperforms the base LLM, for example: \emph{``What flags should I use to select clean galaxy samples from the \gls{DP1} Object table?''}.
\item Queries that probe known failure modes, such as questions about very recent events, highly specific technical details, or information that exists only in private systems, for example: \emph{``What is the magnitudeError column in the \gls{DP1} Object table?''} (a non-existent column) or \emph{``Can I use the \gls{DP0}.2 tutorials to work with \gls{DP1} data?''} (version confusion).
\end{enumerate}
\noindent
For each query, we define either an expected answer or an expected behavior (e.g., declining to answer or requesting clarification).
Together, the three types assess whether the system handles general knowledge correctly, whether retrieval meaningfully improves responses on Rubin-specific questions, and whether the system fails gracefully when it lacks the information to answer.
Running each query with and without retrieval enabled quantifies the improvement contributed by the \gls{RAG} component.
 
\subsection{RAGAS Framework}
\label{sec:ragas}
 
For automated evaluation, we use the \href{https://docs.ragas.io/}{RAGAS} framework \cite{Es2023}, an open-source toolkit that provides reference-free metrics specifically designed for RAG systems.
RAGAS evaluates system quality along four axes: \emph{faithfulness}, \emph{answer relevancy}, \emph{context precision}, and \emph{context recall}.
Together these cover both the retrieval and generation stages of the pipeline, allowing failures to be attributed to the appropriate stage.
 RAGAS also supports the definition of custom metrics, enabling domain-specific evaluation criteria to be added alongside the built-in ones.
For Rubin, this opens the possibility of metrics that verify whether a response correctly identifies the relevant data product, uses valid catalog column names, or produces syntactically correct \gls{ADQL} queries---aspects of correctness that general-purpose metrics cannot assess.
 
\subsection{Current performance and diagnosis}
\label{sec:performance}
 
Initial qualitative evaluation indicates that response quality does not yet meet the standard required for community deployment.
While the core pipeline is functional, preliminary assessment suggests that degradation arises from multiple interacting sources across the full pipeline rather than any single bottleneck; systematic measurement using the RAGAS framework against the validation query set is planned.
 
On the \emph{retrieval} side, the current fixed-size chunking strategy produces chunks that frequently split relevant context across boundaries, reducing context recall; the small-to-big and semantic chunking strategies described in Section~\ref{sec:chunking} are being evaluated as alternatives.
The inherent challenges of astrophysics and observatory retrieval, including vocabulary variation, context-dependent terminology, and dense technical language, also contribute; improvements here are being pursued through chunking strategy, hybrid retrieval, and corpus quality.
 
On the \emph{generation} side, systematic prompt engineering has not yet been performed.
The current system prompt and retrieval-augmented prompt template were constructed heuristically; structured optimization is planned.
We are also evaluating alternative LLM backends beyond \texttt{gpt-3.5-turbo}, as more recent models offer stronger reasoning over retrieved context, better code generation capabilities, and more reliable adherence to system prompt instructions.
 
The composition and coverage of the ingested corpus itself is also under review.
Gaps in source coverage, inconsistent document quality, and stale content all reduce the system's ability to ground responses in authoritative information.
Duplication of content across sources can saturate the retrieval results with redundant chunks, displacing more relevant content from the top-$k$ results and, when different versions are out of sync, introducing contradictory context that degrades response quality.
 
\section{Future Directions}
\label{sec:future}
 
The current prototype demonstrates the viability of the \gls{RAG} approach for Rubin Observatory and has established the core infrastructure and deployment tooling.
The following sections describe planned near-term improvements and the longer-term vision for the system.
 
\subsection{Near-term improvements}
\label{sec:nearterm}
 
\begin{itemize}

\item\textbf{Chunking strategy optimization.}
The current fixed-size chunking is a primary source of retrieval degradation, producing chunks that split relevant context across boundaries and reduce recall. Improved chunking strategies (Section~\ref{sec:chunking}) are expected to increase retrieval precision, ensure that retrieved chunks contain complete and coherent answers, and preserve executable code blocks intact for the coding assistance use case.

\item\textbf{Hybrid retrieval.}
Deploying Weaviate's native hybrid search to combine dense vector similarity with sparse BM25 keyword scoring, as described in Section~\ref{sec:query}, to improve recall for queries that include exact identifiers such as column names, API calls, or error messages that pure semantic search may miss.

\item\textbf{Metadata-enriched embedding.}
Prepending document context (source title, section header, document type) to each chunk before embedding, so that the vector representation encodes provenance alongside content, improving retrieval precision for queries where the source or section of a document is relevant.

\item\textbf{Prompt engineering.}
Systematic optimization of the system prompt and retrieval-augmented prompt template to steer the model toward more accurate, well-structured, and appropriately scoped responses.
The system prompt governs how the LLM interprets retrieved context, when it declines to answer, and how it attributes its sources; careful tuning of these instructions has a significant impact on response quality.
 
\item\textbf{Model benchmarking.}
Evaluating alternative LLM backends, including more recent OpenAI models, Anthropic's Claude, and open-source alternatives, to optimize the trade-off between response quality, latency, and cost.
Adopting a different LLM may also require a corresponding change of embedding model: Anthropic recommends VoyageAI embeddings for use with Claude, and VoyageAI models are strong on scientific and technical documents, code, and the preservation of semantic relationships in complex, specialized terminology, making them a natural candidate if the system moves to a Claude-based backend.
 
\item\textbf{Corpus expansion and automation.}
Expanding the ingested corpus to cover all documentation relevant to each data release and automating the ingestion pipeline with update triggers linked to source platforms (GitHub webhooks, API polling for Jira and Confluence, Slack channel ingestion).
 
\item\textbf{Productizing.}
Transitioning from prototype to a production-quality service with robust monitoring, logging, usage analytics, and automatic re-ingestion of updated and new data sources.

\end{itemize}

\subsection{Towards agentic RAG}
\label{sec:agentic_rag}
 
The current prototype follows a fixed pipeline: each query passes through retrieval, augmentation, and generation in a single pass without the ability to iterate, reformulate queries, or decompose complex multi-step questions.
This limits effectiveness on queries that require reasoning across multiple sources or iterative refinement of a response.
 
Agentic \gls{RAG} represents a natural evolution by introducing an explicit control layer that coordinates how the system reasons over external context.
Rather than following a fixed pipeline, an agentic system can dynamically decide which collections to search, validate retrieved information for consistency, reformulate queries when initial retrieval is insufficient, invoke external tools, and iteratively refine responses.
 
These capabilities are well suited to the most ambitious Rubin use cases described in Section~\ref{sec:motivation}: schema-aware query construction that verifies column names against the current catalog schema; tutorial-guided science workflows that assemble multi-step code examples from different sources; and cross-platform synthesis that combines information from technotes, Jira tickets, and forum posts into a coherent answer.
We plan to investigate agentic \gls{RAG} architectures as the system matures beyond the current prototype.

\section{Summary}
\label{sec:summary}
 
We have presented the design and implementation of a \gls{RAG}-based virtual assistant for information discovery at the Vera C. Rubin Observatory, deployed as a service on the \gls{RSP}.
The system addresses a real operational need: enabling both Rubin  Observatory staff and the Rubin science community to efficiently locate accurate information across a large, fragmented, and continuously evolving documentation ecosystem.
The prototype is functional and deployed on the RSP at both the cloud-based US Data Access Center and the on-premise USDF at SLAC, and is being prepared to support the community during the Early Science period.
As the \gls{LSST} survey progresses and both data volume and documentation grow across successive releases, the importance of this capability will only increase.
Rather than a static documentation virtual assistant, a value proposition that has eroded as general-purpose LLMs have become web-connected, the longer-term vision is a science enablement assistant that helps users go from a science question to working code against real Rubin data, grounded in authoritative and current sources.
The \gls{RAG} value is not in knowing what the documentation says; it is in knowing which specific API call, column name, or query pattern will actually work on the current data release.